\documentclass[apj]{emulateapj}
\usepackage{graphicx}

\begin{document}
\shorttitle{Parallel Spectrum}

\title{On the parallel spectrum in MHD Turbulence}
\author{Andrey Beresnyak}
\affil{Los Alamos National Laboratory, Los Alamos, NM, 87545}
\affil{Nordita, KTH Royal Institute of Technology and Stockholm University, SE-10691 Stockholm, Sweden}

\bibliographystyle{apj}

\begin{abstract}
  Anisotropy of MHD turbulence has been studied extensively for many years,
  most prominently by measurements in the solar wind and high resolution simulations.
  The spectrum parallel to the local magnetic field was observed to be steeper
  than perpendicular spectrum, typically $k^{-2}$, consistent with the widely accepted
  \citet{GS95} model. In this Letter I looked deeper into the nature
  of the relation between parallel and perpendicular spectra and argue that this $k^{-2}$   
  scaling has the same origin as the $\omega^{-2}$ scaling of Lagrangian frequency   
  spectrum in strong hydrodynamic turbulence. This follows from the fact that Alfv\'en 
  waves propagate along magnetic field lines. It is now became clear that the observed
  anisotropy can be argued without invocation of the ``critical balance'' argument
  and is more robust that was previously thought. The relation between parallel  
  (Lagrangian)    
  and perpendicular (Eulerian) spectra is inevitable consequence of strong turbulence of 
  Alfv\'en waves, rather than a conjecture based on the uncertainty relation.
  I tested this using high-resolution simulations of MHD turbulence, in particular I 
  verified that the cutoff of the parallel spectrum scales as Kolmogorov timescale, not 
  lengthscale.
\end{abstract}


\maketitle
\section{Introduction} 

Astrophysical and space plasmas are well-conductive, can be described as a magnetohydrodynamic fluid, which is usually turbulent \citep{armstrong1995,biskamp2003,BL14}.
At the same time high quality measurements in the solar wind has been available
for more than two decades \citep{goldstein1995}.
The presence of large scale magnetic field is expected to change the dynamics dramatically. Analytic weak turbulence theory \citep{galtier2000}
found that turbulent cascade proceeds in the direction perpendicular
to the mean field, resulting in stronger turbulence, which should eventually break down weak turbulence
on sufficiently small scales.
 Similar qualitative arguments earlier lead \citet{GS95}
to suggest that the inertial range of MHD turbulence will be strong turbulence. They also argued that such turbulence will be ``critically balanced'',
or marginally strong, with the linear propagation term always contributing comparably with the nonlinear interaction term predicting the $k_\| \sim k_\perp^{2/3}$ anisotropic cascade, which has found support in numerics, e.g., \citet{cho2000,maron2001}. 
The $k_{\perp}^{-5/3}$ perpendicular spectrum and $k_\| \sim k_\perp^{2/3}$ 
anisotropy results is a $k_\|^{-2}$ parallel spectrum.  One observation is paramount for our understanding of this parallel spectrum. While \citet{GS95} suggested a closure
model predicting the $k_\| \sim k_\perp^{2/3}$ anisotropy in the frame associated with the global mean field, it was not observed
in \citet{cho2000}, rather this anisotropy was observed in the structure function measurement performed in the frame associated
with the local magnetic field. Similarly, \citet{horbury2008} observed the $k_\|^{-2}$ parallel scaling using the wavelet technique
and associating parallel direction to the direction of the local field.

Given the importance of the parallel spectrum for a variety of phenomena, e.g.
resonant scattering of solar energetic particles, the measurement of the parallel
spectrum in the solar wind attracted considerable attention, see, e.g., \citet{horbury2008,podesta2009,osman2009,wicks2010,luo2010}. These measurements
followed the prescription of the local field direction and generally confirmed
the $k_\|^{-2}$ scaling, however, the debate surrounding 
the critical balance argument and the nature of anisotropy continues 
\citet{Grappin2010,Grappin2013}.

In this Letter I will argue that there is a conceptually simpler way
to look at the MHD anisotropy, namely as a relation between Lagrangian and Eulerian
spectra. I will also introduce the statistically averaged one-dimensional spectrum
along the field line and show that high resolution numerics 
support steep parallel spectra, consistent with $k_\|^{-2}$, just like
in the solar wind measurements. 

\section{Strong turbulence and Lagrangian spectrum}

Strong turbulence was suggested to be scale-local and self-similar in \citet{kolm41},
which lead to his $k^{-5/3}$ Eulerian power spectrum of velocity perturbations.
Another basic spectrum of hydrodynamic turbulence is called Lagrangian frequency
spectrum, which statistically evaluates how the velocity of the fluid element changes with time.
Assuming that the dot product of the total time derivative of the velocity and the velocity vector itself is a work done
upon a fluid element, one could estimate $\delta \bf{v}_\tau \cdot \delta \bf{v}_\tau/\tau$ as the turbulence energy cascade
rate per unit mass $\epsilon$, measured in ${\rm cm}^2/{\rm s}^3$, also known
as dissipation rate.
More precisely, in stationary turbulence the second-order structure function of velocity should satisfy:
\begin{equation}
SF(\tau)=\langle ({\bf v}(t+\tau)-{\bf v}(t))^2 \rangle \approx \epsilon \tau
\end{equation}
in the inertial range, where ${\bf v}(t)$ is a velocity of a given fluid element. Such time structure function is dual to the frequency spectrum of $E(\omega)=\epsilon \omega ^{-2}$ \citep{ll06a,corrsin1963,tennekes1972}.
The cutoff of this spectrum is associated with the timescale of critically viscously damped eddies, the Kolmogorov
timescale $\tau_\eta=(\nu/\epsilon)^{1/2}$,
 which has a different dependence on the Reynolds number ${\rm Re}=vL/\nu$ compared
to Kolmogorov length scale $\eta=(\nu^3/\epsilon)^{1/4}$, the cutoff of
the Eulerian spectrum.

Magnetized dynamics is qualitatively different from hydrodynamics in that locally
there is always a propagating wave characteristic. In particular, following a
fluid element, we may find oscillations associated with the wave-train that propagates
though this fluid element in the direction of the local mean magnetic field, which
makes classic Lagrangian measurement of limited value.
The Lagrangian evolution in MHD takes on a different meaning, therefore. The Alfv\'en perturbations
can be decomposed into Els\"asser components ${\bf w^\pm=v\pm B}/\sqrt{4\pi \rho}$,
each of which propagate either along or against the local field direction, i.e. along
the magnetic field line. The functional dependence of such perturbations will take the form $f(s \mp v_A t)$ in the absence of interaction, where $s$ is a distance along the field line. If the nonlinear interaction is present, the trajectory $s =\pm v_A t$
would act as an analogy to hydrodynamic fluid element trajectory if we want to study
Lagrangian dynamics in MHD. 

The above argument suggests that following evolution of ${\bf w^+}$ and
${\bf w^-}$ along the field line in fixed time and in the direction positive in $s$ would be equivalent to following
evolution of ${\bf w^+}$ backward in time and ${\bf w^-}$ forward in time. This simple
argument has been already fruitful in explaining the asymmetric Richardson diffusion of magnetic field lines \citep{B13b}. As far as frequency spectrum goes, the sign
of time is unimportant and any measurement of power spectrum along the field line of
either ${\bf v, B}$ or ${\bf w^\pm}$ will be analogous to Lagrangian frequency
spectrum with frequency $f$ replaced by the wavenumber $f/v_A$: 
$E(k_\|) \sim \epsilon v_A^{-1} k_\|^{-2} $. The spatial structure function
will be expressed correspondingly as $SF_\|(l)=\epsilon l v_A^{-1}$.

Another way to argue the $k_\|^{-2}$ parallel scaling is the dimensional
argument using the Alfv\'en symmetry of reduced MHD used in \citet{B12a}.
Indeed, this symmetry dictates that changing $v_A$ while keeping $k_\| v_A$ constant leave equations unchanged.
Therefore, one must keep energy $E(k_\|) dk_\|$ constant under such transformation, which require that $E(k_\|)\sim v_A^{-1}$.
Using scale-locality, i.e., assuming that the spectrum can only depend on $v_A, \epsilon$ and $k_\|$, I  arrive at
\begin{equation}
 E(k_\|) = C_\| \epsilon v_A^{-1} k_\|^{-2}, \label{par_spec}
\end{equation}
where $C_\|$ is dimensionless constant. Logically, this dimensional
argument is a restatement of the Lagrangian spectrum argument. Note that the parallel second-order
spectrum scales linearly with the dissipation rate $\epsilon$, similarly to the {\it third-order} Eulerian scaling and not to $\epsilon^{2/3}$ scaling of the second-order Eulerian spectrum.

Unlike Reduced MHD, full MHD have no exact Alfv\'en symmetry. The arguments in favor of using it in the inertial range are still quite compelling \citep{B12a}. It is interesting to check
if the parallel spectrum still follow Eq.~(\ref{par_spec}) not only
in Alfv\'enic MHD, but in the general MHD case. Especially interesting is the case with zero mean
magnetic field where the $v_A$ will be determined only by local
fluctuations.

\begin{table}
\begin{center}
\caption{Three-dimensional MHD and RMHD simulations}
  \begin{tabular*}{1.00\columnwidth}{@{\extracolsep{\fill}}c c c c c c c}
    \hline\hline
Run  & $N^3$ & Dissipation & $v_A$ &   $\epsilon$ & $\eta$ & $v_A \tau_\eta$ \\
\hline
MHD1 & $1536^3$ & $-5\cdot10^{-10}k^4$   &0.73& 0.091 &  0.0021 & 0.026     \\
MHD2 & $1536^3$ & $-6.2\cdot10^{-10}k^4$ &1.53& 0.728  &  0.0018 & 0.025 \\
   \hline
M1 & $1024^3$ & $-1.75\cdot10^{-4}k^2$   &1& 0.06 &  0.0031 &  0.044 \\
M2 & $2048^3$ & $-7\cdot10^{-5}k^2$       &1& 0.06  &  0.00155 & 0.028 \\
M3 & $4096^3$ & $-2.78\cdot10^{-5}k^2$   &1& 0.06 &  0.00077 & 0.017 \\
\hline
M1H & $1024^3$ & $-1.6\cdot10^{-9}k^4$   &1& 0.06 &  0.0030 &  0.045                  \\
M2H & $2048^3$ & $-1.6\cdot10^{-10}k^4$  &1& 0.06  &  0.00152 &  0.029 \\
M3H & $4096^3$ & $-1.6\cdot10^{-11}k^4$ &1& 0.06 &  0.00076 & 0.018  \\
   \hline
\end{tabular*}
  \label{experiments}
\end{center}
\end{table}

\section{Numerics}
First half of the numerical data is from my DNS of strong reduced MHD turbulence
\citep{B14a}, which are well-resolved statistically stationary
driven simulations intended to precisely calculate averaged quantities. Note that
reduced MHD, i.e. Alfv\'en dynamics, does not depend on plasma pressure and
can be applied to situations with different values of plasma $\beta$, from zero
to infinity. I list the most important parameters of these simulations in code units in Table~1 under rows M1-3 and M1H-3H. The only difference between M1-3 and M1H-3H was that the latter were performed with higher order diffusivities. Additionally, I performed simulations of statistically isotropic driven incompressible MHD turbulence with zero mean field with parameters presented in Table~1 rows MHD1-2. For all cases I have calculated the spectra along the magnetic field line, and for the reduced MHD cases I additionally have calculated the one-dimensional spectra along the $x$ direction which was the global mean field direction. 

Three dimensional numerics have modest Re and are 
 are always affected by the finite Re effects. I used a rigorous scaling study method, fairly common in the analysis of experimental data and DNS \citep{sreenivasan1995, gotoh2002, kaneda2003, B12b, B14a}, which compares spectra from simulations with
 several different Re values on the same plot with dimensionless axes.
The parallel spectrum was plotted vs dimensionless wavenumber $k v_A \tau_\eta$ and
compensated by $k^2\epsilon^{-1} v_A$ to see how the scaling is consistent with (\ref{par_spec}). This measurement is presented on Fig~1. For the reduced MHD case the spectra collapsed on the dissipation scale, corresponding to an overall scaling of $k^{-2}$.

\begin{figure}[t]
\begin{center}
\includegraphics[width=1.0\columnwidth]{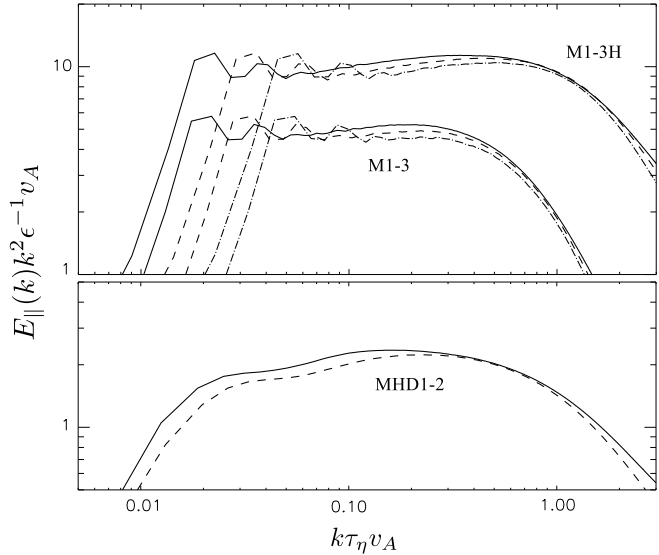}
\end{center}
\caption{Energy spectrum along the magnetic field line compensated by the theoretical scaling $\epsilon k_\|^{-2}$ (\ref{par_spec}).
Solid, dashed and dash-dotted are spectra from $4096^3$,  $2048^3$ and $1024^3$ simulation correspondingly on the upper plot.
The M1-3H has been multiplied by a factor of two to separate the curves.
On the lower plot dashed and solid are MHD1 and MHD2 correspondingly.}
\label{par_sp}
\end{figure}

Given that reduced MHD has precise
Alfven symmetry and the requirement of turbulence to be strong on the outer scale assumes a certain value of $\epsilon$, it does not allow us to check the linear scaling with $\epsilon$ in Eq.~(\ref{par_spec}), as I couldn't vary $\epsilon$ in M1-3.
I used statistically isotropic MHD simulations with zero mean field MHD1-2,
for which Alfv\'en symmetry is absent and the inertial range scaling
(\ref{par_spec}) can not be rigorously argued based on units. 
Despite that, the standard argumentation, introduced by \citet{iroshnikov, kraichnan} is that the RMS magnetic field can play the role of the local mean field and this could
still be regarded as the {\it strong mean field} limit.
I will conjecture that the parallel spectrum will still follow Eq.~(\ref{par_spec}) in the inertial range in this case as well. In the MHD case I used simulations with different $\epsilon$ and substituted the 
RMS field instead of $v_A$ in Eq.~(\ref{par_spec}). Fig.~1 demonstrates that there's an inertial range convergence to $k^{-2}$ even in this zero mean field case. The linear
scaling with $\epsilon$, not $\epsilon^{2/3}$ is also confirmed. 

Another possible spectral measurement is with respect to the global mean field. We do not
expect such scalings to deviate significantly from the perpendicular scalings for the following reason: Alfv\'en waves propagate along the local field direction which
deviates by an angle of $\delta B_L/B_0$ from ${\bf B_0}$, while the angular anisotropy
in this frame is $\delta B_l/B_0$, with inertial range values of $\delta B_l$ much smaller
than the outer scale value of $\delta B_L$. It follows that the anisotropy will be washed
out. 
Fig.~2 presents a measurement of spectrum along the $x$ -- global mean field direction. It is grossly consistent with $-5/3$, i.e. the perpendicular spectral scaling observed in \citet{B14a}.

\section{Discussion}
Critical balance refers to the interaction parameter $\xi=\delta v \lambda_\|/v_A\lambda_\perp$ being around
unity in strong MHD turbulence. It has been first argued based on the uncertainty relation between the wave
frequency and the cascade timescale in \citet{GS95} and had been restated in various forms, including the
decorrelation argument by Gruzinov \citep{maron2001}. While the plausibility arguments like this are certainly useful in qualitative understanding, their apparent generality is problematic. E.g. the decorrelation argument does not explicitly refer to nonlinear interaction, however, it could not be generally valid, as pure propagating solutions with $\xi \gg 1$, strong Alfv\'enic waves, do exist. The naive application of the uncertainty relation argument fails, e.g., in imbalanced turbulence, where it predicts that the anisotropy of the stronger Els\"asser component should be higher than the anisotropy of the weaker component, while
in reality the opposite is true \citep{BL08, BL09a}.
The new argument,
presented in this Letter, circumvents this problem by noticing that the energy cascade is manifested both
in space and time domains, also the parallel direction is equivalent to the time domain. Therefore the well-known anisotropy relation $k_\| \sim k_\perp^{2/3}$ is simply the correspondence between space domain (Eulerian) and frequency domain (Lagrangian) spectra. The old arguments required that the average $\xi$
must be close to unity, while the new argument only requires that the average $\xi$ is a dimensionless, scale-independent
quantity, i.e. a constant, similar to Kolmogorov constant.

Most observational data from the solar wind has been pointing to the $k^{-2}$ parallel spectrum. For example \citet{horbury2008} used a wavelet technique
to follow the local field direction and obtained $k^{-2}$. This has been further improved in \citet{wicks2010} and compared with the global Fourier spectra.
\citet{podesta2009} obtained similar results with wavelets and demonstrated scale-dependent anisotropy. The structure function measurement in \citet{luo2010}
again confirmed the same scaling. Multi-spacecraft measurements allowing better coverage of k-space \citep{osman2009} also confirmed $k^{-2}$.
Earlier measurements in the {\it global} frame, e.g. \citet{matthaeus1990} reported {\it scale-independent} anisotropy, which, as I argued above, 
is consistent with theory and numerics as well. As far as numerics go, the measurements along the local field direction gave the $k^{-2}$
slope, see, e.g., \citet{cho2000,maron2001,BL09a,BL09b,Chen2011,B12b}, while the measurements in the global frame gave scale-independent anisotropy, see, e.g. \citet{Grappin2010} or my Fig~2.
The robustness of the critical balance with properly defined nonlinear timescale has been recently discussed in \citet{Mallet2014}.

\begin{figure}[t]
\begin{center}
\includegraphics[width=1.0\columnwidth]{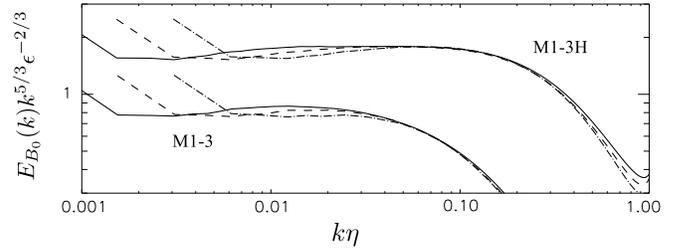}
\end{center}
\caption{The spectra along the global mean field in M1-3, M1-3H. The M1-3H spectra has been multiplied by a factor of two. This plot demonstrates that this energy spectrum scaling is consistent with $-5/3$, i.e. the same as the perpendicular scaling. }
\label{b0_sp}
\end{figure}

Recently, the debate on the parallel spectrum has been revived, in particular \citet{Turner2012} measured quasi-isotropic spectrum in the solar wind after filtering out discontinuities,
while \citet{Grappin2012,Grappin2013} suggested a new model with ``ricochet'' cascade that effectively fills parallel direction and results in the same slope,
as in perpendicular direction, citing \citet{Grappin2010,Turner2012} as motivation. My 
numerical data strongly disfavors this model, as the observed $-2.0 \div -1.9$ parallel
spectral slope is much steeper than either $-5/3$ or the $-1.5$ suggested in \citet{Grappin2012,Grappin2013}. It is also not clear why global measurement \citep{Grappin2010}
should support the alternative model or whether filtering in \citet{Turner2012} interfere with the local field direction enough to destroy the weaker $k^{-2}$ parallel spectrum. 

Measurements of Lagrangian frequency spectrum in hydrodynamics has been performed my many authors, see, e.g., \cite{yeung2006} and references therein,
and showed correspondence with the theoretical $\omega^{-2}$. 
First direct measurement of Lagrangian frequency spectrum in statistically isotropic MHD turbulence
has been performed in \cite{Busse2010} and tentatively confirmed the $\omega^{-2}$ scaling, but the results
from simulations with strong mean field were less clear.
The connection between Lagrangian frequency spectrum and parallel spacial spectrum
has not been made in \cite{Busse2010}, also as I argued above, the classic Lagrangian
measurement could be meaningless in MHD as high-frequency perturbations cross
the fluid element causing oscillatory velocity changes not associated with the energy cascade. In this Letter I argued that the measurement of the spectrum {\it along the magnetic field line} is similar to the measurement of Lagrangian frequency spectrum
and therefore the \citet{GS95} scale-dependent anisotropy is a simple relation between Eulerian and Lagrangian spectra.

\begin{figure}
\begin{center}
\includegraphics[width=1.0\columnwidth]{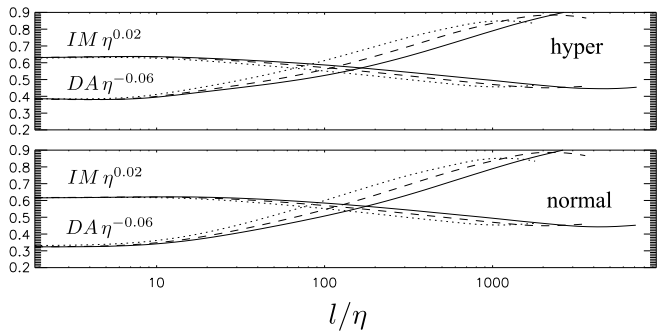}
\end{center}
\caption{Scaling study of alignment measures $DA=\langle
|\delta {\bf v} {\bf \times} \delta {\bf b} |\rangle /\langle |\delta v \delta
b|\rangle$ and $IM=\langle |\delta (w^+)^2- \delta (w^-)^2|\rangle /\langle \delta
(w^+)^2+ \delta (w^-)^2\rangle$ from M1-3H (top) and M1-3 (bottom). The alignment slopes converge to relatively small values,
e.g., 0.06 for $DA$ which is smaller than 0.25, predicted by \citet{Boldyrev2006}. See also \citet{BL09b,B11,B12b}.}
\label{align_conv}
\end{figure}

Using scaling study argument I found that the best convergence correspond to
$k^{-2}$, however the deviation around $0.1$ in scaling exponent on medium scales is evident on Fig.~1 and 
is somewhat interesting from theoretical viewpoint as a long-range finite-Re effect.
It is clear that $-1.9$ slope on the medium scales on Fig.~1 is still much steeper than Eulerian $-5/3\approx -1.7$, but what is the nature of this deviation? Firstly, the prediction of the models with so-called dynamic
alignment modifies only perpendicular spectrum, leaving parallel spectrum unchanged,
at the expense of higher anisotropy \citep{Boldyrev2005,Boldyrev2006}. Secondly, even
if such modification was suggested by some inertial-range theory, it would be inconsistent
with my numerics, as the $0.1$ correction is not universal and disappear with higher Re measurements.

In the past several years various spectral scalings, deviations from theoretically predicted scalings and alignment measures has been studied in some detail \citep{BL09b,B11,B12a}. The overall picture seems to be that while moderate Re shows scale-dependency
of several alignment measures, normally in the range of $0.1-0.2$, in the higher Re
measurements these alignment measures flatten out and their slopes are fairly close to zero, see, e.g., Fig.~3. Similarly, the deviation from the expected perpendicular $-1.7$ slope is around $\sim 0.2$ in the medium
scales, but disappeared when higher resolution data became available. The $\sim 0.1$ deviation
of the parallel slope fits nicely into this tendency. We see that the deviations
from theoretical scalings had been observed so far only within around an order
of magnitude in scale from the driving scale and modifications of theory of the inertial range, such as \citet{Chandran2014} are probably excessive at this point. Further
solar wind measurements with better statistics and/or larger scale numerics will
help to shed light on this problem.

\section{Acknowledgements}
The author was supported though the DOE LDRD program, while the computer time was provided by the DOE INCITE program and the NSF XSEDE grant TG-AST110057. This research
used resources of the ALCF at Argonne National Laboratory, which is supported by the DOE under contract DE-AC02-06CH11357. The author also thanks support and hospitality of Nordita.


\def\apj{{\rm ApJ }}           
\def\apjl{{\rm ApJ }}          
\def\apjs{{\rm ApJ }}          
\def\grl{{\rm GRL }}
\def\aap{{\rm A\&A } }
\def\mnras{{\rm MNRAS } }
\def\physrep{{\rm Phys. Rep. } }               
\def\prl{{\rm Phys. Rev. Lett. }} 
\def\pre{{\rm Phys. Rev. E }} 
\def\jgr{{\rm JGR }}

\bibliography{all}

\begin{thebibliography}{42}
\expandafter\ifx\csname natexlab\endcsname\relax\def\natexlab#1{#1}\fi

\bibitem[{{Armstrong} {et~al.}(1995){Armstrong}, {Rickett}, \&
  {Spangler}}]{armstrong1995}
{Armstrong}, J.~W., {Rickett}, B.~J., \& {Spangler}, S.~R. 1995, \apj, 443, 209

\bibitem[{{Beresnyak}(2011)}]{B11}
{Beresnyak}, A. 2011, \prl, 106, 075001

\bibitem[{{Beresnyak}(2012{\natexlab{a}})}]{B12b}
---. 2012{\natexlab{a}}, \mnras, 422, 3495

\bibitem[{{Beresnyak}(2012{\natexlab{b}})}]{B12a}
---. 2012{\natexlab{b}}, \prl, 108, 035002

\bibitem[{{Beresnyak}(2013)}]{B13b}
---. 2013, \apjl, 767, L39

\bibitem[{{Beresnyak}(2014)}]{B14a}
---. 2014, \apjl, 784, L20

\bibitem[{{Beresnyak} \& {Lazarian}(2008)}]{BL08}
{Beresnyak}, A., \& {Lazarian}, A. 2008, \apj, 682, 1070

\bibitem[{{Beresnyak} \& {Lazarian}(2009{\natexlab{a}})}]{BL09b}
---. 2009{\natexlab{a}}, \apj, 702, 1190

\bibitem[{{Beresnyak} \& {Lazarian}(2009{\natexlab{b}})}]{BL09a}
---. 2009{\natexlab{b}}, \apj, 702, 460

\bibitem[{{Beresnyak} \& {Lazarian}(2014)}]{BL14}
---. 2014, Magnetic Fields in Diffuse Media (Springer), 163--226

\bibitem[{{Biskamp}(2003)}]{biskamp2003}
{Biskamp}, D. 2003, {Magnetohydrodynamic Turbulence} (Cambridge: Cambridge
  University Press)

\bibitem[{{Boldyrev}(2005)}]{Boldyrev2005}
{Boldyrev}, S. 2005, \apjl, 626, L37

\bibitem[{{Boldyrev}(2006)}]{Boldyrev2006}
---. 2006, \prl, 96, 115002

\bibitem[{{Busse} {et~al.}(2010){Busse}, {M{\"u}ller}, \&
  {Gogoberidze}}]{Busse2010}
{Busse}, A., {M{\"u}ller}, W.-C., \& {Gogoberidze}, G. 2010, Physical Review
  Letters, 105, 235005

\bibitem[{{Chandran} {et~al.}(2014){Chandran}, {Schekochihin}, \&
  {Mallet}}]{Chandran2014}
{Chandran}, B.~D.~G., {Schekochihin}, A.~A., \& {Mallet}, A. 2014,
  ArXiv:1403.6354

\bibitem[{{Chen} {et~al.}(2011){Chen}, {Mallet}, {Yousef}, {Schekochihin}, \&
  {Horbury}}]{Chen2011}
{Chen}, C.~H.~K., {Mallet}, A., {Yousef}, T.~A., {Schekochihin}, A.~A., \&
  {Horbury}, T.~S. 2011, \mnras, 415, 3219

\bibitem[{{Cho} \& {Vishniac}(2000)}]{cho2000}
{Cho}, J., \& {Vishniac}, E.~T. 2000, \apj, 539, 273

\bibitem[{{Corrsin}(1963)}]{corrsin1963}
{Corrsin}, S. 1963, Journal of Atmospheric Sciences, 20, 115

\bibitem[{{Galtier} {et~al.}(2000){Galtier}, {Nazarenko}, {Newell}, \&
  {Pouquet}}]{galtier2000}
{Galtier}, S., {Nazarenko}, S.~V., {Newell}, A.~C., \& {Pouquet}, A. 2000,
  Journal of Plasma Physics, 63, 447

\bibitem[{{Goldreich} \& {Sridhar}(1995)}]{GS95}
{Goldreich}, P., \& {Sridhar}, S. 1995, \apj, 438, 763

\bibitem[{{Goldstein} {et~al.}(1995){Goldstein}, {Smith}, {Balogh}, {Horbury},
  {Goldstein}, \& {Roberts}}]{goldstein1995}
{Goldstein}, B.~E., {Smith}, E.~J., {Balogh}, A., {Horbury}, T.~S.,
  {Goldstein}, M.~L., \& {Roberts}, D.~A. 1995, \grl, 22, 3393

\bibitem[{Gotoh {et~al.}(2002)Gotoh, Fukayama, \& Nakano}]{gotoh2002}
Gotoh, T., Fukayama, D., \& Nakano, T. 2002, Physics of Fluids, 14, 1065

\bibitem[{{Grappin} \& {M{\"u}ller}(2010)}]{Grappin2010}
{Grappin}, R., \& {M{\"u}ller}, W.-C. 2010, \pre, 82, 026406

\bibitem[{{Grappin} {et~al.}(2012){Grappin}, {M{\"u}ller}, \&
  {G{\"u}rcan}}]{Grappin2012}
{Grappin}, R., {M{\"u}ller}, W.-C., \& {G{\"u}rcan}, {\"O}. 2012,
  ArXiv:1209.4450

\bibitem[{{Grappin} {et~al.}(2013){Grappin}, {M{\"u}ller}, {Verdini}, \&
  {G{\"u}rcan}}]{Grappin2013}
{Grappin}, R., {M{\"u}ller}, W.-C., {Verdini}, A., \& {G{\"u}rcan}, {\"O}.
  2013, ArXiv:1312.3459

\bibitem[{{Horbury} {et~al.}(2008){Horbury}, {Forman}, \&
  {Oughton}}]{horbury2008}
{Horbury}, T.~S., {Forman}, M., \& {Oughton}, S. 2008, Physical Review Letters,
  101, 175005

\bibitem[{Iroshnikov(1964)}]{iroshnikov}
Iroshnikov, P. 1964, Soviet Astronomy, 7, 566

\bibitem[{{Kaneda} {et~al.}(2003){Kaneda}, {Ishihara}, {Yokokawa}, {Itakura},
  \& {Uno}}]{kaneda2003}
{Kaneda}, Y., {Ishihara}, T., {Yokokawa}, M., {Itakura}, K., \& {Uno}, A. 2003,
  Physics of Fluids, 15, L21

\bibitem[{{Kolmogorov}(1941)}]{kolm41}
{Kolmogorov}, A. 1941, Akademiia Nauk SSSR Doklady, 30, 301

\bibitem[{Kraichnan(1965)}]{kraichnan}
Kraichnan, R. 1965, Physics of Fluids, 8, 1385

\bibitem[{{Landau} \& {Lifshitz}(1944)}]{ll06a}
{Landau}, L.~D., \& {Lifshitz}, E.~M. 1944, {Fluid mechanics} (Moscow: Moscow)

\bibitem[{{Luo} \& {Wu}(2010)}]{luo2010}
{Luo}, Q.~Y., \& {Wu}, D.~J. 2010, \apjl, 714, L138

\bibitem[{{Mallet} {et~al.}(2014){Mallet}, {Schekochihin}, \&
  {Chandran}}]{Mallet2014}
{Mallet}, A., {Schekochihin}, A.~A., \& {Chandran}, B.~D.~G. 2014,
  ArXiv:1406.5658

\bibitem[{{Maron} \& {Goldreich}(2001)}]{maron2001}
{Maron}, J., \& {Goldreich}, P. 2001, \apj, 554, 1175

\bibitem[{{Matthaeus} {et~al.}(1990){Matthaeus}, {Goldstein}, \&
  {Roberts}}]{matthaeus1990}
{Matthaeus}, W.~H., {Goldstein}, M.~L., \& {Roberts}, D.~A. 1990, \jgr, 95,
  20673

\bibitem[{{Osman} \& {Horbury}(2009)}]{osman2009}
{Osman}, K.~T., \& {Horbury}, T.~S. 2009, Annales Geophysicae, 27, 3019

\bibitem[{{Podesta}(2009)}]{podesta2009}
{Podesta}, J.~J. 2009, \apj, 698, 986

\bibitem[{Sreenivasan(1995)}]{sreenivasan1995}
Sreenivasan, K. 1995, Physics of Fluids, 7, 2778

\bibitem[{{Tennekes} \& {Lumley}(1972)}]{tennekes1972}
{Tennekes}, H., \& {Lumley}, J.~L. 1972, {First Course in Turbulence}
  (Cambridge, MA: MIT Press)

\bibitem[{{Turner} {et~al.}(2012){Turner}, {Chapman}, \&
  {Gogoberidze}}]{Turner2012}
{Turner}, A.~J., {Chapman}, S.~C., \& {Gogoberidze}, G. 2012, ArXiv:1210.1695

\bibitem[{{Wicks} {et~al.}(2010){Wicks}, {Horbury}, {Chen}, \&
  {Schekochihin}}]{wicks2010}
{Wicks}, R.~T., {Horbury}, T.~S., {Chen}, C.~H.~K., \& {Schekochihin}, A.~A.
  2010, \mnras, 407, L31

\bibitem[{{Yeung} {et~al.}(2006){Yeung}, {Pope}, \& {Sawford}}]{yeung2006}
{Yeung}, P.~K., {Pope}, S.~B., \& {Sawford}, B.~L. 2006, Journal of Turbulence,
  7, 58

\end{thebibliography}

\end{document}